\documentstyle[12pt,epsfig]{elsart}

\begin{document}
\begin{frontmatter}

\title{Influence of topology on\\
the performance of a neural network}
\author{J.J. Torres\corauthref{cor1}
},
\corauth[cor1]{Corresponding author}
\ead[url]{jtorres@onsager.ugr.es}
\author{M.A. Mu\~{n}oz, J. Marro, and P.L. Garrido }

\address{Department of Electr. and Matter Physics, and 
Institute Carlos I for Theoretical and Computational Physics,
University of Granada, E-18071 Granada, Spain.}

\date{}

\begin{abstract}
We studied the computational properties of an attractor neural network (ANN)
with different network topologies. Though fully connected neural networks
exhibit, in general, a good performance, they are biologically unrealistic, as
it is unlikely that natural evolution leads to such a large connectivity. 
We demonstrate that, at finite temperature, the capacity
to store and retrieve binary patterns is higher for ANN with scale--free
(SF) topology than for highly random--diluted Hopfield networks with
the same number of synapses. We also show that, at zero temperature, the
relative performance of the SF network increases with increasing values of
the distribution power-law exponent. Some consequences and possible applications of our findings are discussed.
\end{abstract}

\end{frontmatter}
\section{Introduction}
There is a growing interest in evolving complex networks, in
particular, networks with scale--free (SF) topology~\cite{ABreview02,Mendes,VS03}. SF occurs in many different contexts, including the WWW and the Internet, 
e-mail and scientific--citation networks, ecological, protein and gene 
interaction networks, etc. In these examples, the degree $k$ of a vertex, 
i.e., the number of arcs linking it to other vertices, is power-law 
distributed, $ P(k)\sim k^{-\gamma }$ (see figure~\ref{fig1}). This implies 
that the network includes a relatively large number of nodes with small 
connectivity, defining what we call the network {\em boundary}, and a few 
nodes, the {\em hubs}, with a large connectivity, comparable to the network 
size $N$. As a consequence, SF networks exhibit the interesting 
{\em small-world} property, that is, the average path length between two 
nodes is very small compared to the network size. 

Evolving networks with such complex topology are also common in biology. 
Neuronal networks, for instance, seem to exhibit the
small-world property. This was recently demonstrated in a set of 
{\em in-vitro} experiments of growing cultured neurons~\cite{SGSA02}.
Although an impressive amount of work has been done in the last few years
concerning SF networks, it has only recently been reported on the specific
consequences of such an architecture on the performance of auto--associative
neural networks~\cite{SAFA03,GM03}. The authors in~\cite{SAFA03} show that a
SF neural network is able to store and retrieve $P$ patterns with a lower
computer--memory cost than the fully--connected Hopfield neural network.
They also find a similar performance with a (biologically unrealistic)
nearest--neighbor hypercubic Ising lattice. The authors in~\cite{GM03} study
the zero temperature behavior of different topologies, namely, the
Barab\'asi--Albert (BA) SF, small--world and random diluted networks, and a
better performance for the random diluted case than for the other topologies
is reported. However, for the relative large mean connectivity these authors
use ($\langle k\rangle =50$), the BA network has not the SF property~\cite{SAFA03}, so that this result lacks interest.

We here report on the influence of topology on the associative--memory task
of the network as a function of temperature. In particular we focus on two
main issues, namely, on the robustness of the network performance against
thermal noise for varying topology, and on the effect of varying the SF
connectivity distribution $P(k)$ on the network performance.

\section{Definition of Models}
Consider the BA {\it evolving} network~\cite{ABreview02} with $N$ nodes and 
$\eta (N-\eta _{0})$ links. Here, $\eta _{0}$
is the initial number of nodes generating the network, $\eta \leq \eta _{0}$
is the number of links that are added during the {\it evolution} at each
time step, and $N$ is the final number of nodes in the network. This will
latter be generalized to consider other SF networks. In order to have a
neural system with the chosen topology, we place a binary neuron, $s_{i}=1$
or $0,$ at each node $i,$ and then \textquotedblleft store\textquotedblright\ $P$
binary random patterns, ${\bf \xi }^{\mu }\equiv \left\{ \xi _{i}^{\mu }=1%
\mbox{ or }0\right\} ,$ $\mu =1,\ldots ,P,$ with mean activity level $%
\langle \xi _{i}^{\mu }\rangle =1/2.$ This is done in practice by
associating a synaptic intensity $\omega _{ij}$ at each link according to
the Hebbian learning rule, 
\begin{equation}
\omega _{ij}=\frac{1}{N}\sum_{\mu =1}^{P}(2\xi _{i}^{\mu }-1)(2\xi _{j}^{\mu
}-1).
\end{equation}%
\begin{figure}[th]
\centerline{\psfig{file=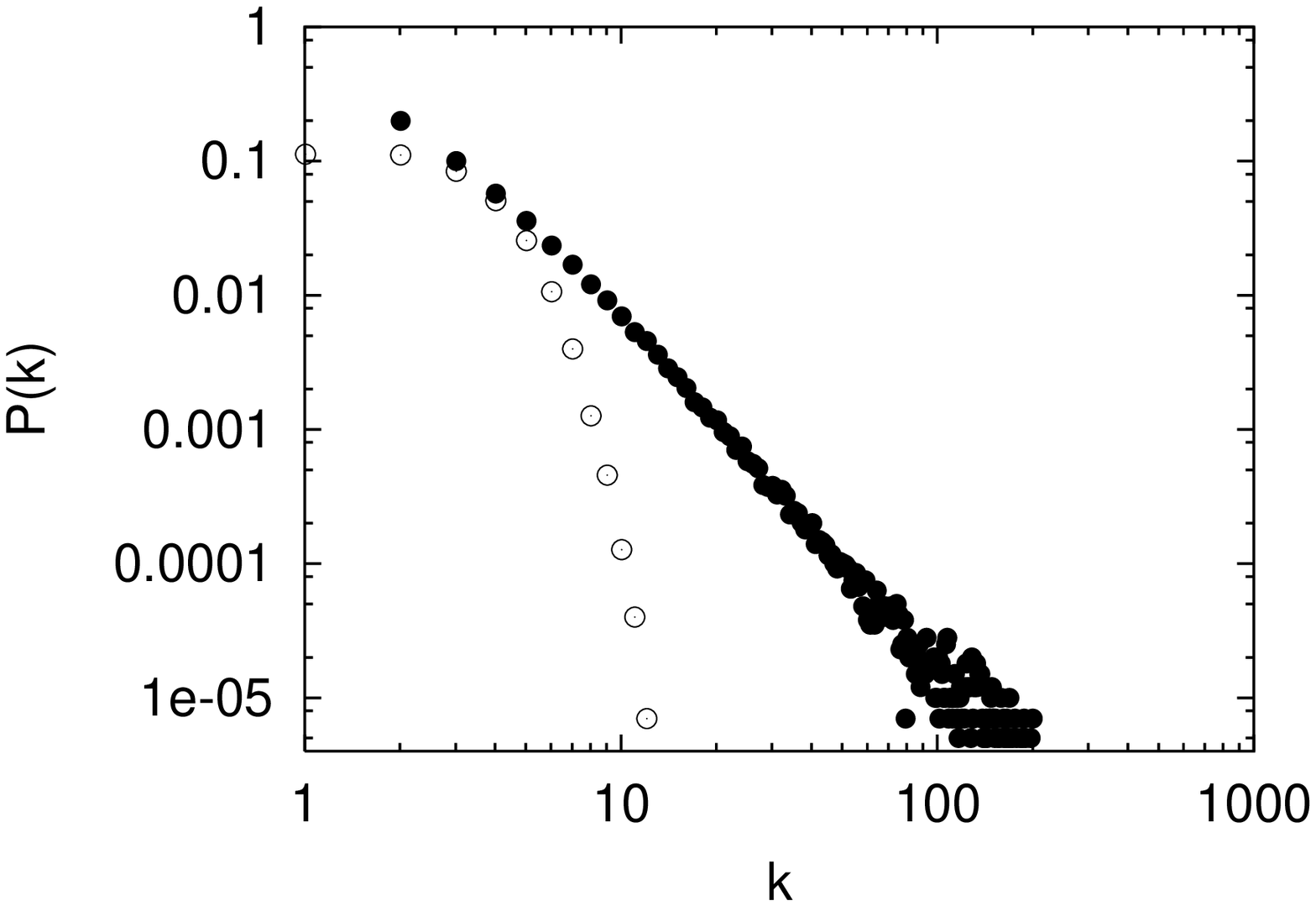,width=7.1cm}
\psfig{file=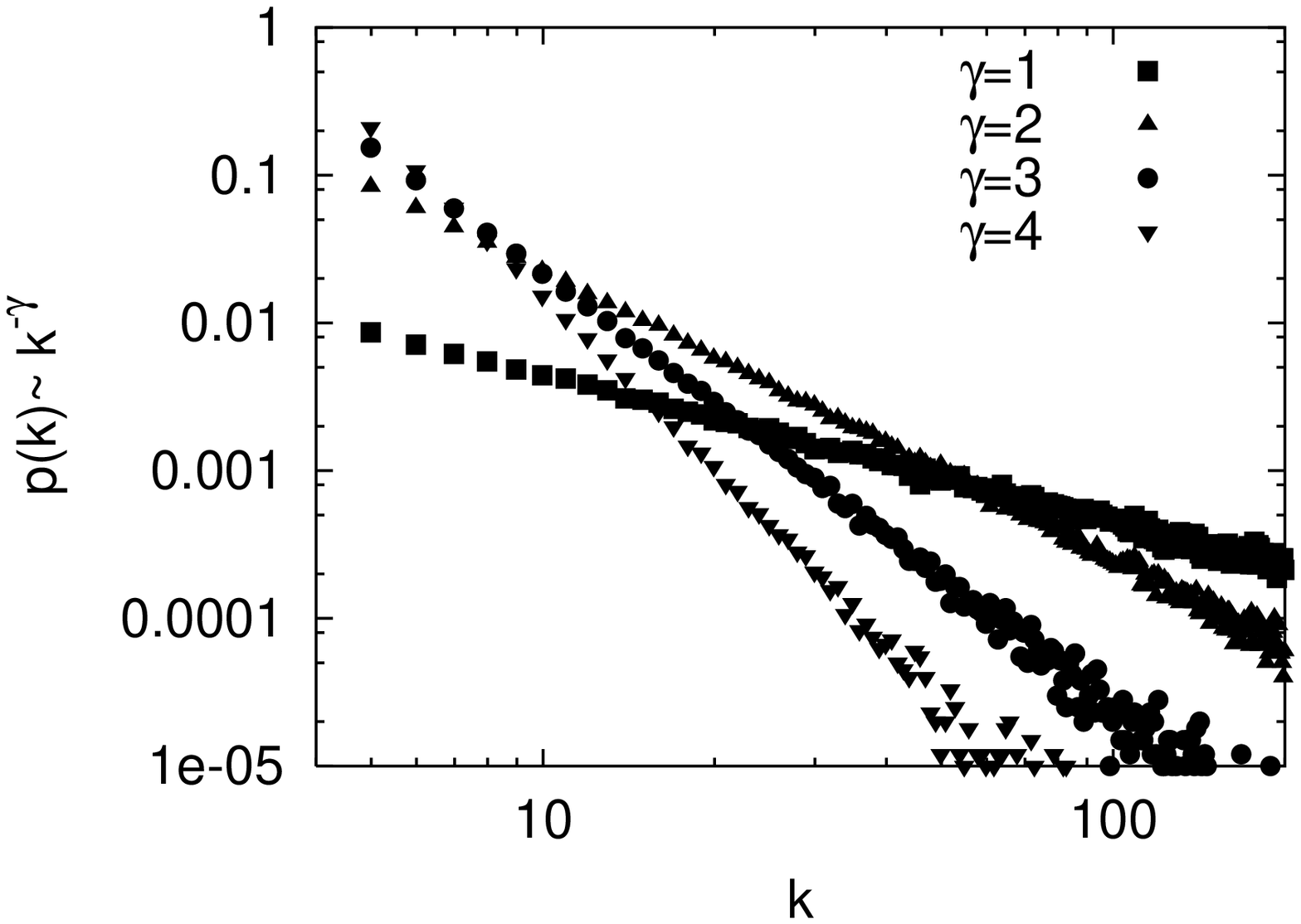, width=7.1cm }}
\caption{Left: Connectivity probability distributions for a SF network with $\protect\gamma =3$ (filled circles) and
for a HDH network (open circles). Right: Connectivity probability distributions generated for a SF network by tuning the exponent $\gamma.$ Each data point corresponds to an average over $100$ networks with $N=4000$ neurons each.}
\label{fig1}
\end{figure}
A meaningful direct comparison of this SF neural network (SFNN) and the
standard Hopfield neural network cannot be made because the second case
involves $N^{2}$ synaptic connections. A hypercubic Ising lattice has the
same number of synapses than the SFNN; however, real neural systems are
known to exhibit more complex neuron connectivity than the Ising network.
Consequently, we compare the performance of the SFNN with that of a {\em %
highly diluted} Hopfield network (HDHN). The HDHN is obtained from the
standard Hopfield network by randomly suppressing synapses until only $\eta
(N-\eta _{0})$ of them remain, i.e., the number of synapses scales as $N$
and not as $N^{2}$. To maintain the SF behavior in the BA network, the value
of $\eta $ must be very small compared to the network size, that is, $\eta
\ll N$~\cite{SAFA03}. The connectivity distribution of the HDHN is illustrated in figure~\ref{fig1} (left). The main differences between this distribution 
and the corresponding one
for a SF network is that the latter has no typical connectivity value. More 
specifically, the SF network distribution is a
power--law while the HDHN distribution has a maximum and an Gaussian
decay and, consequently, may be characterized by a (typical) mean
connectivity.

\begin{figure}[ht]
\centerline{\psfig{file=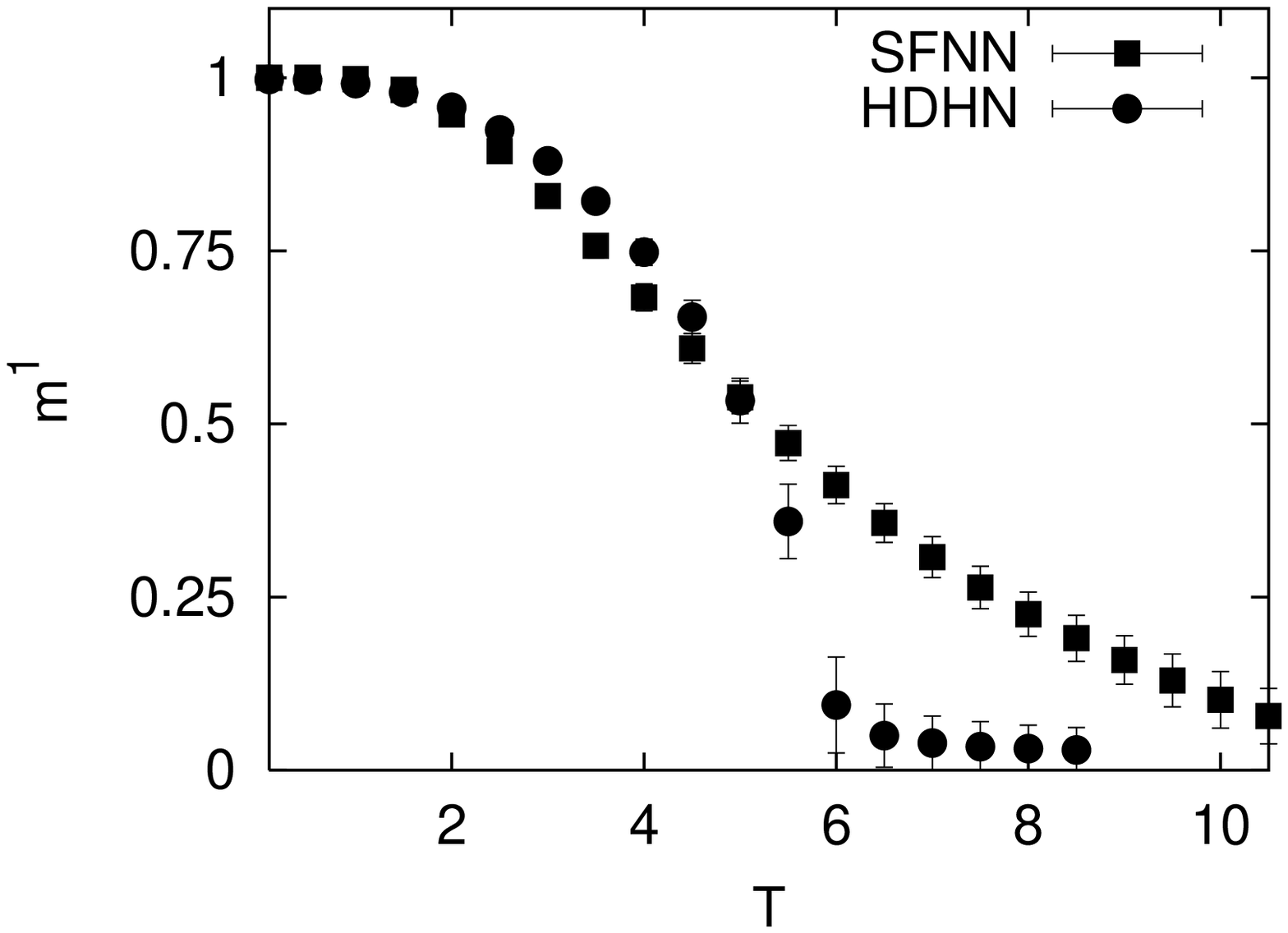,width=7.25cm}
\psfig{file=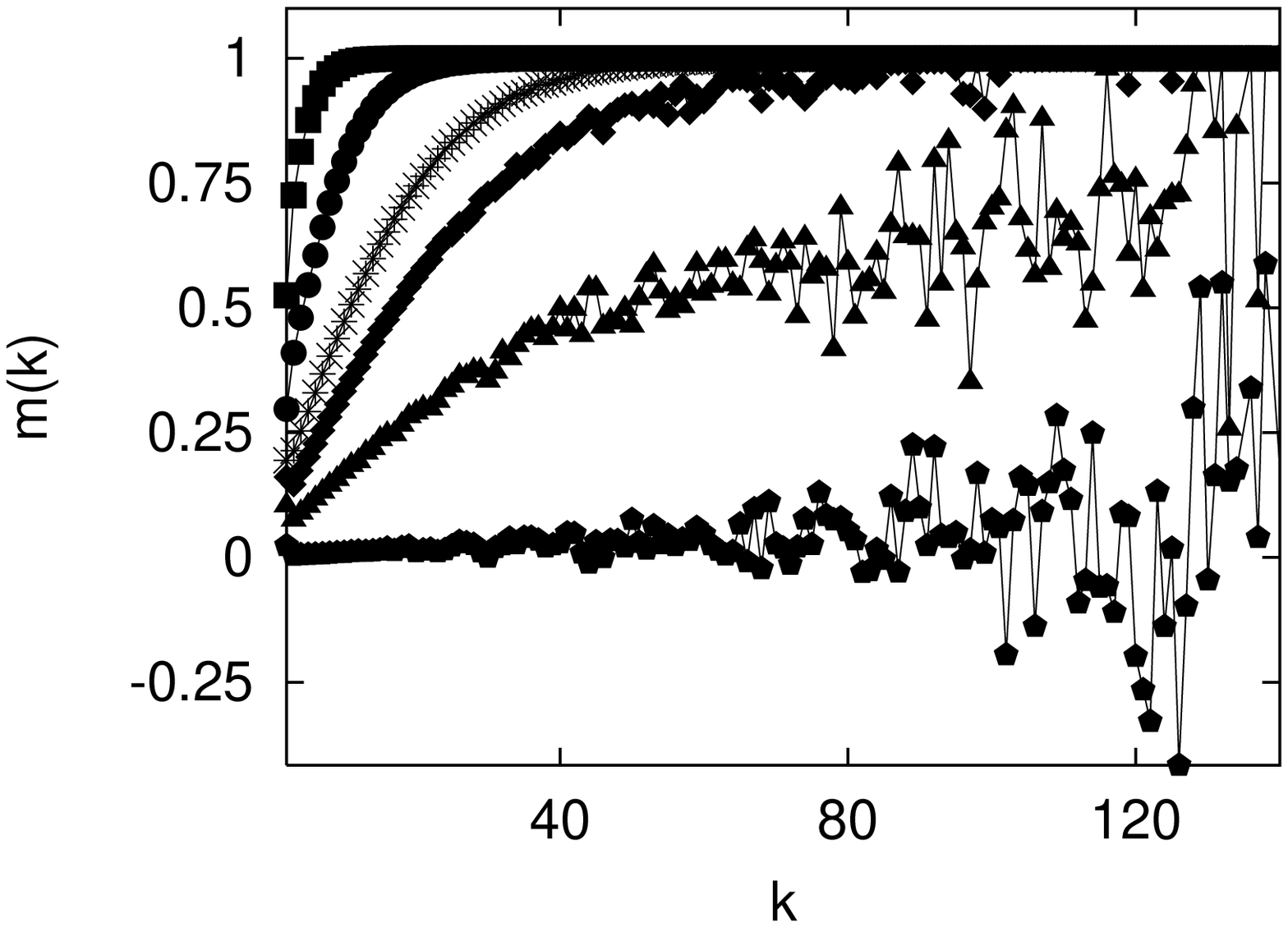,width=7.25cm}
}
\caption{Right: Overlap curves for varying temperature (in arbitrary units) for a BA network (filled squares) and a HDHN (filled circles). Data correspond to an average over $100$
realizations of a network with $N=1600$ neurons, $P=1$ and $\protect\eta =%
\protect\eta _{0}=3.$ Left: Local overlap $m(k)$ for a BA network as a function of the connectivity $k$ for increasing values of the temperature (successive decreasing curves).}
\label{fig2}
\end{figure}

A relevant magnitude to monitor in order to compare the performance of
different topologies is the overlap function, defined for pattern $\nu $ as 
\begin{equation}
m^{\nu }\equiv \frac{2}{N}\sum_{i}(2\xi _{i}^{\nu }-1)s_{i}.
\end{equation}
The performance of the two networks is compared in figure~\ref{fig2} for $%
P=1$ and $\eta =3.$ This clearly shows that, excluding very low temperature,
the retrieval of information as characterized by $m^{\nu }$ is better for
the SFNN than for the HDHN. In both cases, the retrieval of
information deteriorates as $P$ is increased. However, we also observe that,
at finite temperature, the performance of the SFNN increases significantly
if one considers only the retrieval of 
information concerning neurons with a
connectivity degree higher than certain value, $k_{0},$ i.e., the hubs~\cite{TMMG03}. This can be understood on simple grounds by computing the local
overlap $m_{i}^{1}$ with one pattern for neuron $i$ and averaging over all
neurons with the same connectivity $k.$
The resulting mean overlap for a
given connectivity $m(k)$ is plotted in figure~\ref{fig2} (right) for 
different temperatures. 
This shows that, even at high temperature, the local 
overlap for hubs is close to one whereas it is very small for the boundary 
(fluctuations in the lower curves are due to the small number of
hubs present in the network for the relative small network size we are
using). This finding reflects the
``negative'' role of the boundary and the ``positive'' role of the hubs on the 
SFNN performance during each retrieval experiment when thermal fluctuations 
are considered. This observation is in agreement with the $T=0$ behavior
reported in~\cite{GM03}.

Another important issue is how the exponent of the
distribution influences the performance of SF network for associative memory
tasks. In order to analyze this, we studied networks characterized by
different power--law exponents. With this end, we used a Molloy--Red (MR) SF network~\cite{MR95} with $P(k)\sim k^{-\gamma },$
where $\gamma $ is a tunable parameter. As illustrated in figure~\ref{fig1} (right), the number of neurons in the network with a high connectivity increases as $\gamma $ is decreased. 

\begin{figure}[ht]
\centerline{
\psfig{file=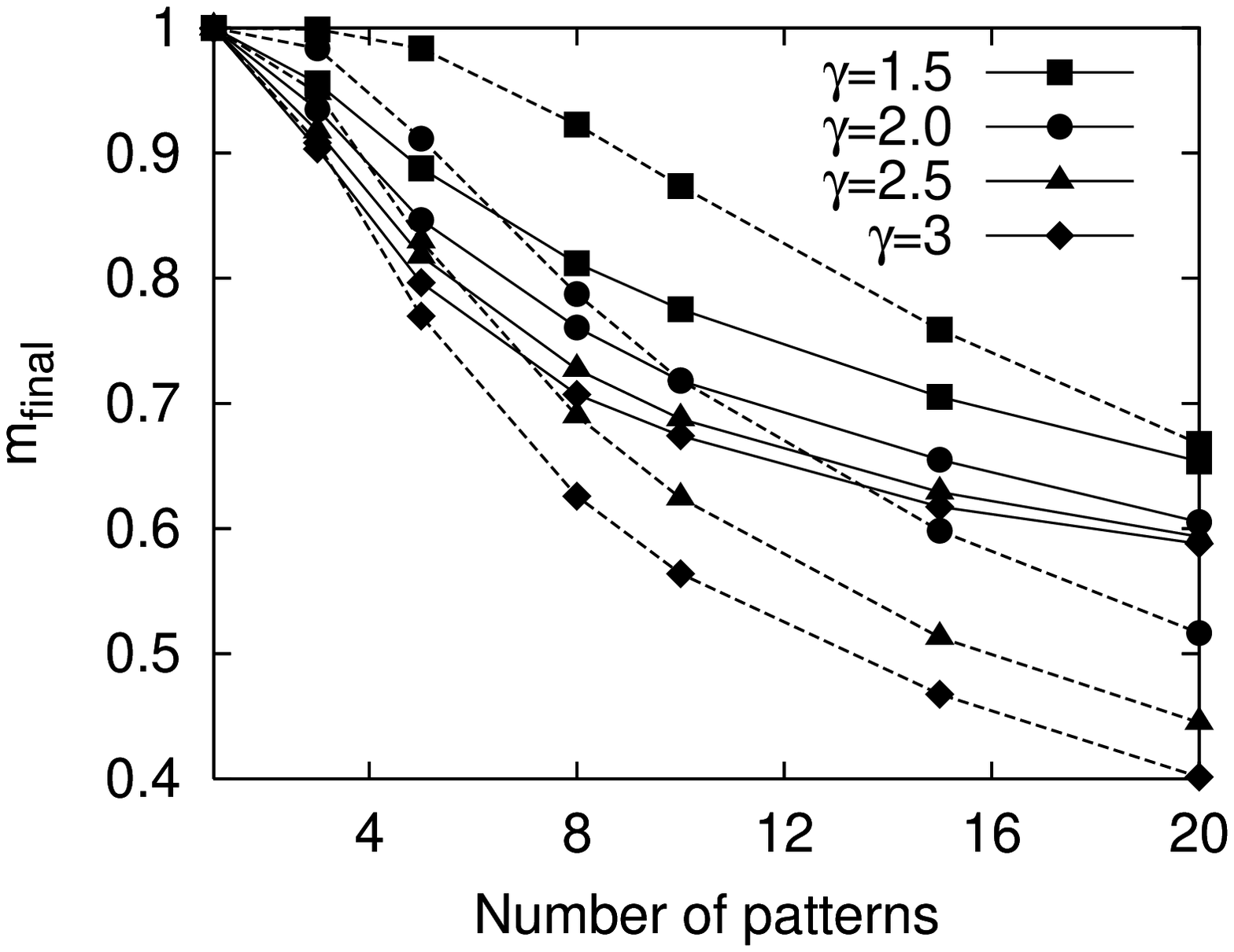,width=7.25cm}\psfig{file=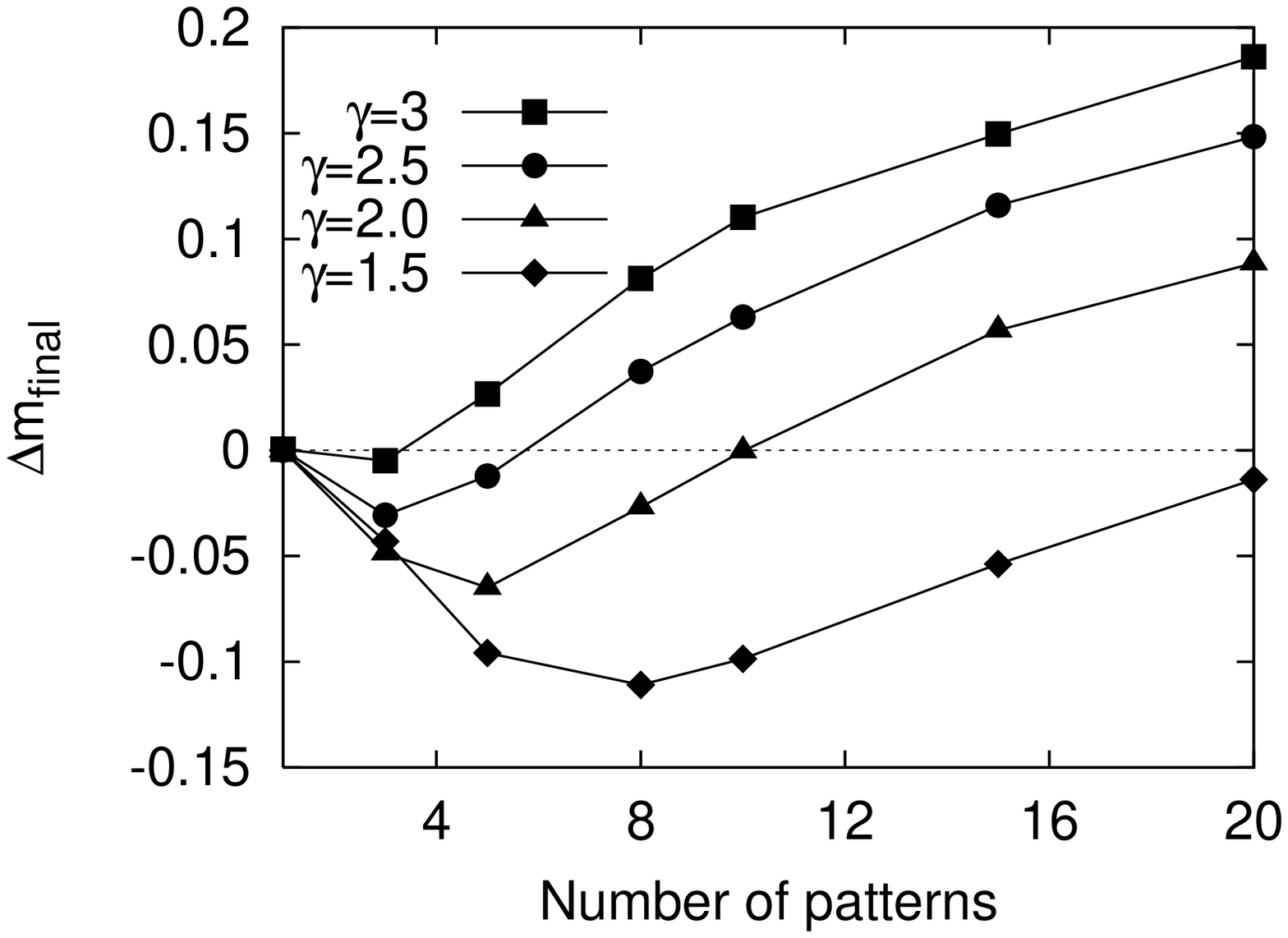,width=7.25cm}}
\caption{Left: Zero temperature performance of a Molloy-Red SFN for $%
\protect\gamma =1.5,2,2.5,3$ (solid lines) compared to the performance of
different HDHN with the same number of synapses (dashed lines), and as a
function of the number of stored patterns $P.$ Right: Relative difference in
the performance for the cases showed in the left panel.}
\label{fig3}
\end{figure}

Even more interesting is when one compares the behavior of the MR network 
with that of the HDHN in the limit $T=0$ (cf. figure~\ref{fig3}). As thermal
fluctuations are then suppressed, the network performance is only perturbed
by the interference among the stored patterns. In order to consider the
limit of interest, we started with one of the stored patterns, and computed
the state of each neuron according to the deterministic rule 
$s_{i}(t+1)=\Theta (h_{i}(t)).$
Here, $\Theta (x)$ is the Heaviside step function, and $h_{i}\equiv
\sum_{j}^{(i)}w_{ij}s_{j}$ is the local field associated with neuron $i$ with
the sum over all neurons $j$ connected to it. At the end of the
network evolution, we recorded the overlap, $m_{final},$
with the starting pattern as a function of the total number of stored
patterns $P$. In order to visualize the difference in performance between
the two types of networks, we defined $\Delta m_{final}\equiv
m_{final}^{SFNN}-m_{final}^{HDHN}$. 
Figure~\ref{fig3} shows how this 
difference in performance varies with the number of stored patterns. 
The graph illustrates that the SFNN has a better and better performance 
as compared with the HDHN as the number of stored patterns is increased. 
This effect is enlarged as $\gamma $ is increased. This can be understood by 
considering the different decays of $P(k)$ for large $k$ in both SFNN and 
HDHN, and the fact that for increasing values of $\gamma ,$ the relative 
position of $\langle k\rangle $ moves to the left for both topologies, but 
due to the power--law decay the effect of the hubs remains for the SFNN.

Summing up, the topology of a neural network has a key
role in the processes of memorization and retrieval of patterns. In
particular, neural networks with scale--free topology may exhibit a better
performance than Hopfield--like networks with the same number of synapses
distributed randomly over the network. Our study can be useful to understand
the role of regions with different connectivity degrees in real neural
systems during memorization and retrieval of information. In particular, it
may improve our understanding of how fluctuations or perturbations in the
typical number of synapses of some brain areas can affect the processing of
information and memorization in these regions. Our study
also suggests the convenience of developing new methods to store the more 
relevant information into the hubs, increasing in this way the effective 
network-performance and efficiency. It would be desirable to check our 
findings against experimental observations on real neural systems focusing 
on the topology which is built up by natural selection.

\section*{Acknowledgments}
We tanks Dr. Pastor-Satorras for fruitful suggestions. This work was supported by the Spanish MCyT and FEDER ``Ram\'on y Cajal''  contract and project no BFM2001-2841.

\end{document}